\documentclass[prd,preprint,showpacs,preprintnumbers,nofootinbib,eqsecnum,superscriptaddress]{revtex4}

 \usepackage[dvips,final]{graphicx}
  \usepackage{amssymb}
   \usepackage{amsmath}
    \usepackage{amsfonts}
     \usepackage{epsfig}
      \usepackage{bm}

\usepackage{mathpazo}

\usepackage[section]{placeins}

\usepackage{multirow}
\usepackage{ctable}
\usepackage{booktabs}
\usepackage{array}
\usepackage{tabularx}
\usepackage{xcolor}
\usepackage{pstricks}

\begin{document}

\title{New mechanisms for double charmed meson production at the LHCb}

\author{Rafa{\l} Maciu{\l}a}
\email{rafal.maciula@ifj.edu.pl} \affiliation{Institute of Nuclear
Physics, Polish Academy of Sciences, Radzikowskiego 152, PL-31-342 Krak{\'o}w, Poland}

\author{Vladimir A. Saleev}
\email{saleev@samsu.ru}\author{Alexandra V. Shipilova}
\email{alexshipilova@samsu.ru}\affiliation{Samara State Aerospace
University, Moscow Highway, 34, 443086, Samara, Russia}

\author{Antoni Szczurek\footnote{also at University of Rzesz\'ow, PL-35-959 Rzesz\'ow, Poland}}
\email{antoni.szczurek@ifj.edu.pl} \affiliation{Institute of Nuclear
Physics, Polish Academy of Sciences, Radzikowskiego 152, PL-31-342 Krak{\'o}w, Poland}

\date{\today}

\begin{abstract}
We discuss production of $D^0 D^0$ (and ${\bar D}^0 {\bar D}^0$) pairs
related to the LHCb Collaboration results for $\sqrt{s}$ = 7 TeV in proton-proton scattering. We
consider double-parton scattering (DPS) mechanisms of double $c \bar
c$ production and subsequent $cc \to D^{0}D^{0}$ hadronization as well as double $g$ and mixed
$g c\bar c $ production with $gg \to D^{0}D^{0}$ and $gc \to D^{0}D^{0}$ hadronization calculated 
with the help of the scale-dependent hadronization functions of Kniehl et al.
Single-parton scattering (SPS) mechanism of digluon production is also taken into account.
We compare our results with several correlation observables in azimuthal angle $\varphi_{D^{0}D^{0}}$
between $D^{0}$ mesons or in dimeson invariant mass $M_{D^{0}D^{0}}$.
The inclusion of new mechanisms with $g \to D^{0}$ fragmentation leads to larger cross sections,
than when including only DPS mechanism $cc \to D^{0}D^{0}$ with standard scale-independent fragmentation functions. Some consequences of the
presence of the new mechanisms are discussed. In particular a larger $\sigma_{eff}$
is needed to describe the LHCb data. There is a signature that $\sigma_{eff}$
may depend on transverse momentum of $c$ quarks and/or $\bar c$ antiquarks.
\end{abstract}


\maketitle

\section{Introduction}

Some time ago two of us predicted that at large energies relevant
for the LHC the production of double charm should be dominated
by the double-parton scattering (DPS) mechanism \cite{Luszczak:2011zp}.
In the first calculation the cross section for each step was calculated
in the leading-order (LO) collinear approach.
However, the LO collinear approach is not sufficient for
a detailed description of actual cross section for the $c \bar c$
production. The double $c \bar c$ production was extended next
to the $k_t$-factorization approach which includes effectively
higher-order QCD effects \cite{Maciula:2013kd,vanHameren:2014ava}.
A relatively good description of the LHCb experimental data \cite{Aaij:2012dz} was achieved for both the total
yield and the dimeson correlation observables.
In these calculations the standard scale-independent Peterson fragmentation function (FF) \cite{Peterson:1982ak} was used.
The single-parton scattering (SPS) $g g \to c \bar c c \bar c$
contribution was discussed carefully in both collinear \cite{vanHameren:2014ava}
and $k_t$-factorization \cite{vanHameren:2015wva} approaches.
Their contribution to the $c \bar c c \bar c$ cross section was found to be rather
small and was not able to describe details of the LHCb
data \cite{Aaij:2012dz}.

Studies of inclusive $D$ meson production at the LHC based on scale-independent
FFs have been done in next-to-leading order (NLO) collinear approach within the FONLL scheme \cite{Cacciari}
as well as in the $k_{t}$-factorization \cite{Maciula:2013wg}. In turn, in Ref.~\cite{Kniehl2012} the calculation was done according to the GM-VFNS NLO collinear scheme together with the several scale-dependent FFs of a parton (gluon, $u,d,s,\bar u, \bar d, \bar s, c, \bar c$) to $D$ mesons proposed by Kniehl et al. \cite{Kniehl:2005de,Kniehl:2006mw}, that undergo DGLAP evolution equations. It has been found that important contribution to inclusive production of $D$ mesons comes from gluon fragmentation (see also Ref.~\cite{Kniehl:2005ej}). Similar calculation were done recently also in the
$k_t$-factorization approach with parton Reggeization hypothesis by two of us \cite{Nefedov:2014qea}. They
have also shown there that the new mechanism constitutes a big fraction of the
cross section for $D$ meson production and a good description of the
inclusive $D$ meson production at the LHC was achieved.

In the present paper we wish to investigate how important is the
gluon fragmentation mechanism for the double $D$-meson production, i.e.
double fragmentation of each of the gluons in the gluon dijets in
SPS production and double fragmentation of each of the gluons in the
gluon jet in DPS production mechanism. Here the gluon and digluon production is
considered in the $k_t$-factorization approach with Reggeized gluons
in the t-channel \cite{Nefedov:2013ywa} via subprocesses $RR\to g$
and $RR\to g g$, where $R$ is the Reggeized gluon. In our analysis
we shall use scale-dependent fragmentation functions of Kneesch-Kniehl-Kramer-Schienbein (KKKS08)
\cite{Kneesch:2007ey} as implemented in the code available on the
Web \cite{webpage}.

\section{A sketch of the theoretical formalism}

\begin{figure}[!h]
\begin{minipage}{0.35\textwidth}
 \centerline{\includegraphics[width=1.0\textwidth]{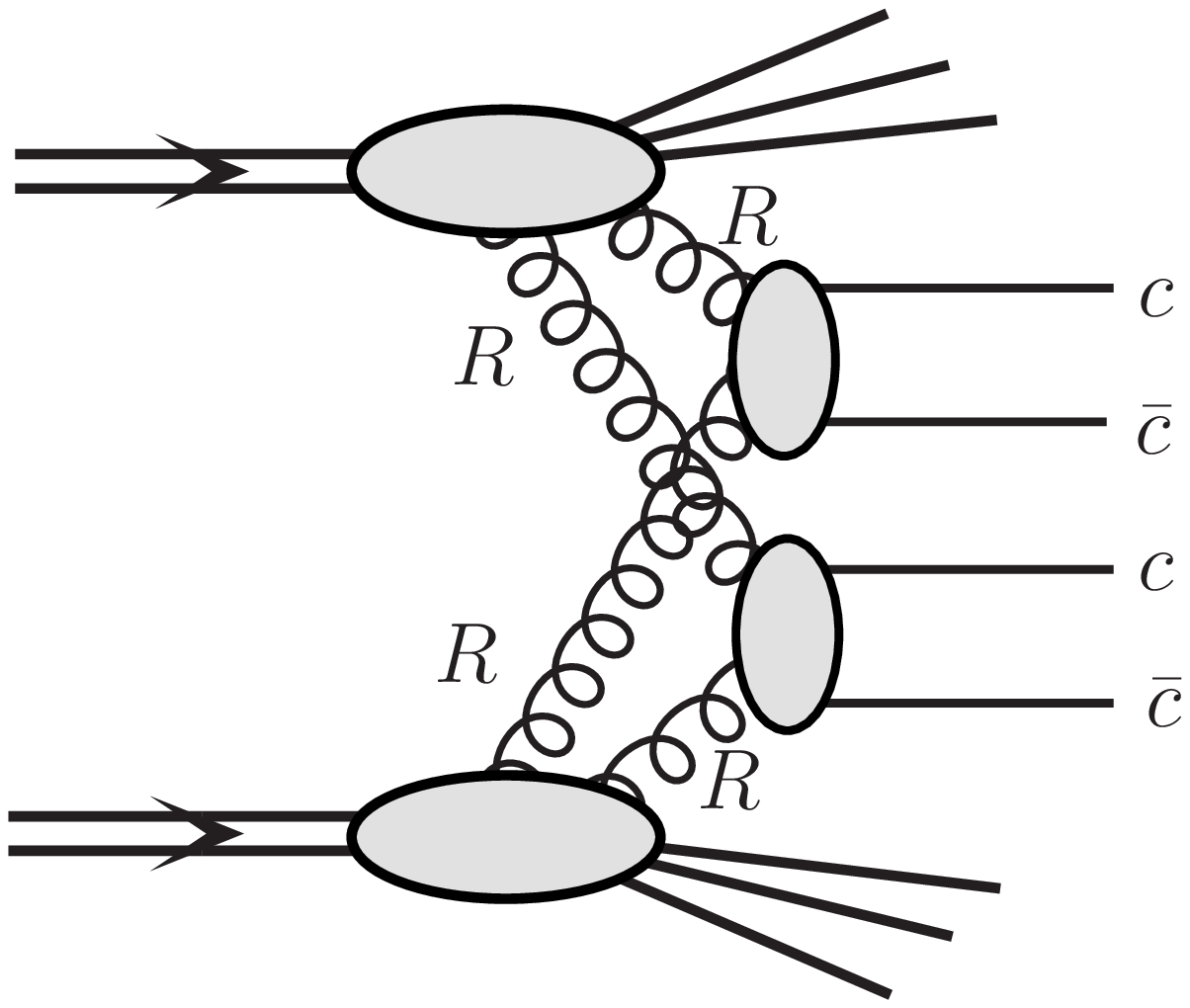}}
\end{minipage}
\hspace{1.5cm}
\begin{minipage}{0.35\textwidth}
 \centerline{\includegraphics[width=1.0\textwidth]{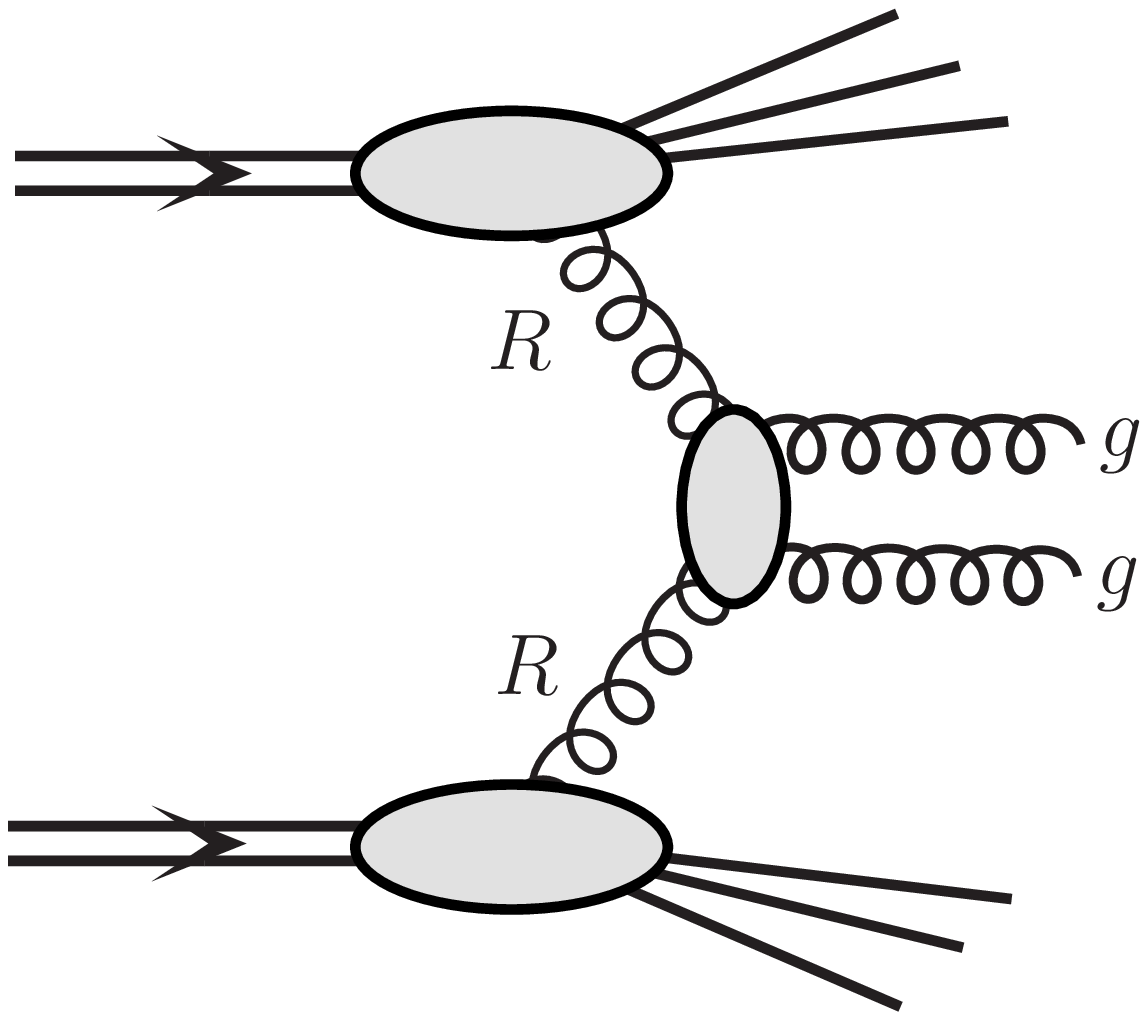}}
\end{minipage}\\
\vspace{0.5cm}
\begin{minipage}{0.35\textwidth}
 \centerline{\includegraphics[width=1.0\textwidth]{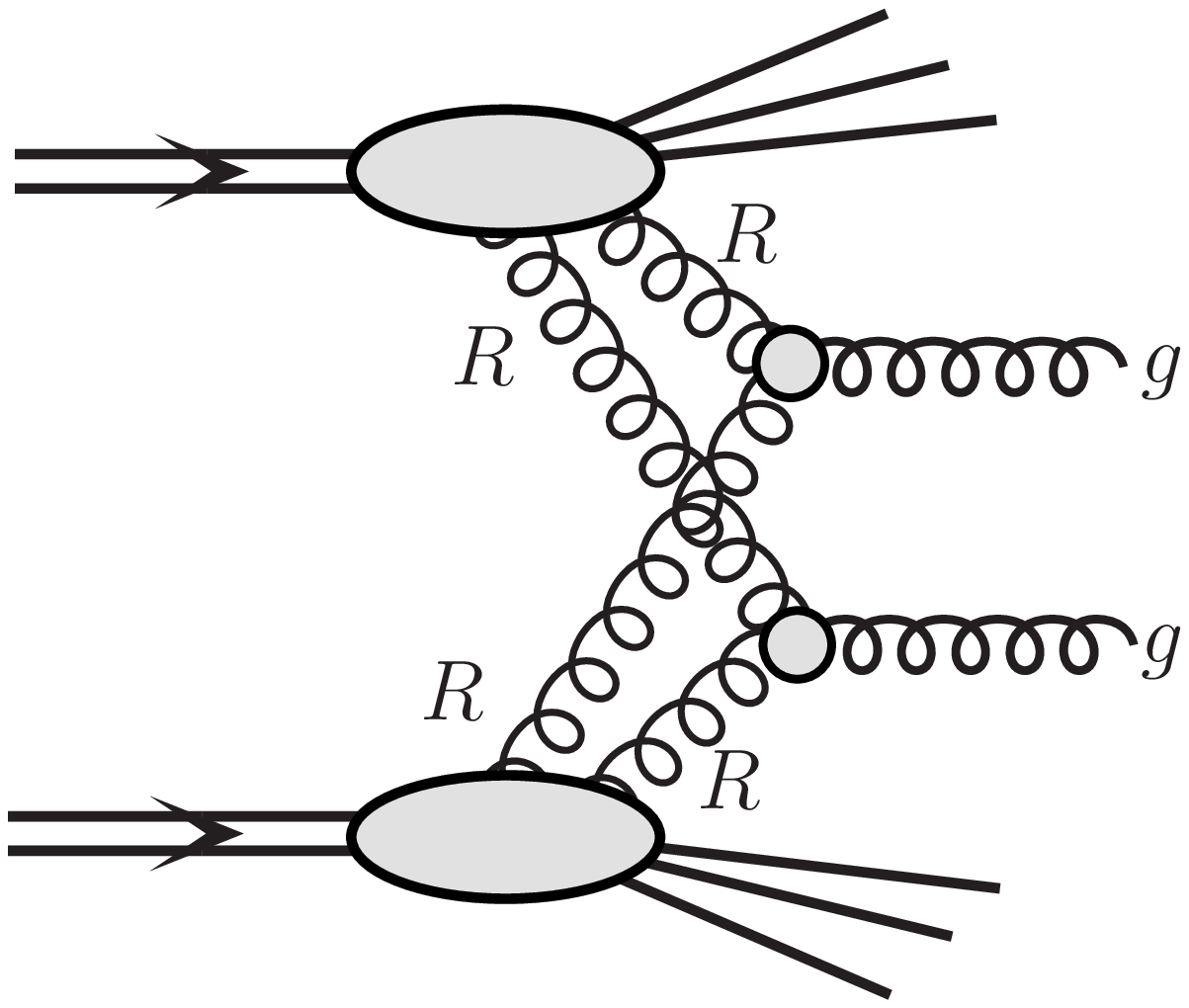}}
\end{minipage}
\hspace{1.5cm}
\begin{minipage}{0.35\textwidth}
 \centerline{\includegraphics[width=1.0\textwidth]{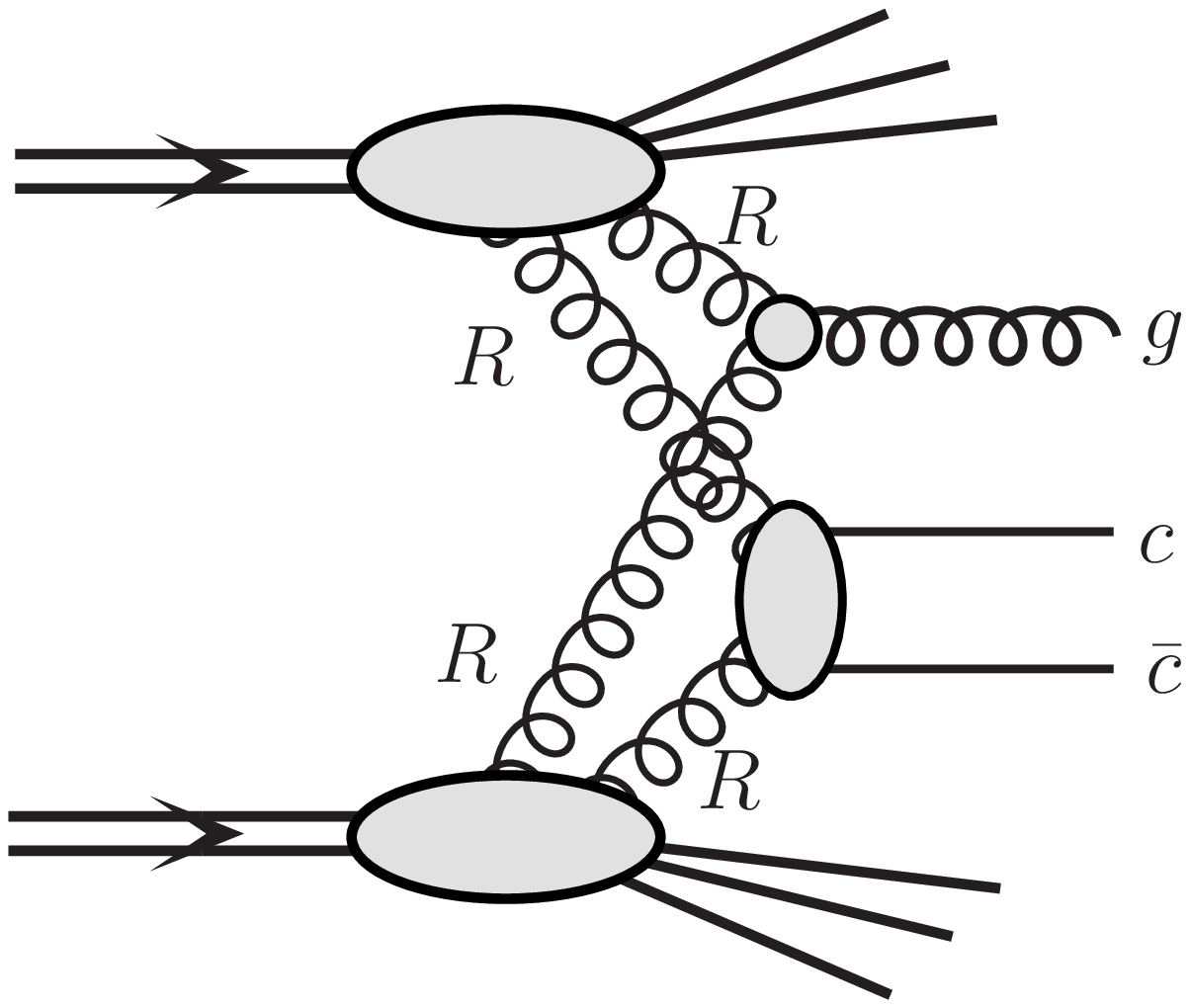}}
\end{minipage}
   \caption{
\small A diagrammatic illustration of the considered mechanisms.
 }
 \label{fig:diagrams}
\end{figure}

We will compare results with first (old) and second (new) approach.
In the second (new) scenario with $g \to D$ fragmentation the number of contributing processes
grows compared to the first (old) scenario with $c \to D$ fragmentation only. 
Naturally a new single-parton scattering mechanism (called
here SPS $gg \to DD$) appears
(top-right panel in Fig.\ref{fig:diagrams}). Since here the two produced gluons are
correlated in azimuth, the mechanism will naturally lead to an
azimuthal correlation between the $D D$ (or $\bar D \bar D$) mesons.
Such a correlation was actually observed in the LHCb experimental data 
\cite{Aaij:2012dz} and could not be explained by the SPS 2 $\to$ 4 perturbative $gg \to c \bar c c \bar c$
contribution (see e.g. Ref.~\cite{vanHameren:2015wva})
which turned out to be rather small.
In the new scenario we have more processes for single $D$ meson
production (two components) and as a consequence many more processes for
the pair production in double-parton scattering.
Now (in the new scenario) there are three classes of DPS contributions.
In addition to the coventional DPS $cc \to DD$ (top-left panel in Fig.\ref{fig:diagrams})
considered in Refs.~\cite{Maciula:2013kd,vanHameren:2014ava,vanHameren:2015wva}
there is a double $g \to D$ (or double $g \to \bar D$) fragmentation mechanism, called here
DPS $gg \to DD$ (bottom-left panel in Fig.\ref{fig:diagrams}) as well as
the mixed DPS $gc \to DD$ contribution (bottom-right panel in Fig.\ref{fig:diagrams}).

DPS cross section for production of $cc$, $gg$ or $gc$ system, assuming factorization of the
DPS model, can be written as:
\begin{eqnarray}
\frac{d \sigma^{DPS}(p p \to c  c  X)}{d y_1 d y_2 d^2
p_{1,t} d^2 p_{2,t}} = 
\frac{1}{2 \sigma_{eff}} \cdot \frac{d \sigma^{SPS}(p p \to c \bar c
X_1)}{d y_1  d^2 p_{1,t}} \cdot \frac{d \sigma^{SPS}(p p \to c \bar
c X_2)}{d y_2 d^2 p_{2,t}}, \label{DPScc_factorization_formula}
\end{eqnarray}
\begin{eqnarray}
\frac{d \sigma^{DPS}(p p \to g g  X)}{d y_1 d y_2 d^2
p_{1,t} d^2 p_{2,t}} =
\frac{1}{2 \sigma_{eff}} \cdot \frac{d \sigma^{SPS}(p p \to g
X_1)}{d y_1  d^2 p_{1,t}} \cdot \frac{d \sigma^{SPS}(p p \to g
X_2)}{d y_2 d^2 p_{2,t}}. \label{DPSgg_factorization_formula}
\end{eqnarray}
\begin{eqnarray}
\frac{d \sigma^{DPS}(p p \to g c  X)}{d y_1 d y_2 d^2
p_{1,t} d^2 p_{2,t}} =
\frac{1}{\sigma_{eff}} \cdot \frac{d \sigma^{SPS}(p p \to g
X_1)}{d y_1  d^2 p_{1,t}} \cdot \frac{d \sigma^{SPS}(p p \to c \bar
cX_2)}{d y_2 d^2 p_{2,t}}. \label{DPSgg_factorization_formula}
\end{eqnarray}
When integrating over kinematical variables one recovers the commonly used pocket formula:
\begin{equation}
\sigma^{DPS}(p p \to c  c  X) = \frac{1}{2 \sigma_{eff}}
\sigma^{SPS}(p p \to c \bar c X_1) \cdot \sigma^{SPS}(p p \to c \bar
c X_2), \label{basic_formula1}
\end{equation}
\begin{equation}
\sigma^{DPS}(p p \to g g  X) = \frac{1}{2 \sigma_{eff}}
\sigma^{SPS}(p p \to g X_1) \cdot \sigma^{SPS}(p p \to g X_2),
\label{basic_formula2}
\end{equation}
\begin{equation}
\sigma^{DPS}(p p \to g c  X) = \frac{1}{\sigma_{eff}}
\sigma^{SPS}(p p \to g X_1) \cdot \sigma^{SPS}(p p \to c\bar c X_2).
\label{basic_formula2}
\end{equation}
The often called pocket-formula is a priori a severe approximation. The flavour, spin and color correlations lead, in principle, to interference effects
that result in its violation as discussed e.g. in Refs.~\cite{Diehl:2011tt,Diehl:2011yj}. Even for unpolarized proton beams, the spin polarization of the two partons from one hadron
can be mutually correlated, especially when the partons are relatively close in phase space (having comparable $x$'s). Moreover, in contrast to the standard single PDFs, the two-parton distributions have a nontrivial color structure which also may lead to a non-negligible correlations effects. 
Such effects are usually not included in phenomenological analyses. They were exceptionally discussed in the context of double charm production \cite{Echevarria:2015ufa}.
However, the effect on e.g. azimuthal correlations between charmed quarks was found there to be very small, much smaller than effects of the SPS
contribution associated with double gluon fragmentation discussed in the present paper.
In addition, including perturbative parton splitting mechanism also leads to a breaking of the pocket-formula \cite{Ryskin:2011kk,Gaunt:2012dd,Gaunt:2014rua}.
This formalism was so far formulated for the collinear leading-order
approach which for charm (double charm) may be a bit academic as this
leads to underestimation of the cross section.
Imposing sum rules also leads to a breaking of the factorized Ansatz
but the effect almost vanishes for small longitudinal momentum fractions 
\cite{Golec-Biernat:2015aza}. Taken the above we will use the pocket-formula in the
following.

In the $k_t$-factorization approach, the cross section for SPS cross
sections can be presented as follows:
\begin{eqnarray}
\frac{d \sigma^{SPS}(p p \to c \bar c X)}{d y_1 d y_2 d^2 p_{1,t} d^2 p_{2,t}}
&& = \frac{1}{16 \pi^2 {(x_1 x_2 S)}^2} \int \frac{d^2 k_{1t}}{\pi} \frac{d^2 k_{2t}}{\pi}
\overline{|{\cal M}_{R R \rightarrow c \bar{c}}|^2} \nonumber \\
&& \times \;\; \delta^2 \left( \vec{k}_{1t} + \vec{k}_{2t} - \vec{p}_{1t} - \vec{p}_{2t}
\right)
{\cal F}(x_1,k_{1t}^2,\mu^2) {\cal F}(x_2,k_{2t}^2,\mu^2),
\label{ccbar_kt_factorization}
\end{eqnarray}
\begin{eqnarray}
\frac{d \sigma^{SPS}(p p \to g g X)}{d y_1 d y_2 d^2 p_{1,t} d^2 p_{2,t}}
&& = \frac{1}{16 \pi^2 {(x_1 x_2 S)}^2} \int \frac{d^2 k_{1t}}{\pi} \frac{d^2 k_{2t}}{\pi}
\overline{|{\cal M}_{R R \rightarrow g g}|^2} \nonumber \\
&& \times \;\; \delta^2 \left( \vec{k}_{1t} + \vec{k}_{2t} - \vec{p}_{1t} - \vec{p}_{2t}
\right)
{\cal F}(x_1,k_{1t}^2,\mu^2) {\cal F}(x_2,k_{2t}^2,\mu^2).
\label{gg_kt_factorization}
\end{eqnarray}

\begin{eqnarray}
\frac{d \sigma^{SPS}(p p \to g X)}{d y d^2 p_{t}}
&& = \frac{\pi}{{(x_1 x_2 S)}^2} \int \frac{d^2 k_{1t}}{\pi} \frac{d^2 k_{2t}}{\pi}
\overline{|{\cal M}_{R R \rightarrow g }|^2} \nonumber \\
&& \times \;\; \delta^2 \left( \vec{k}_{1t} + \vec{k}_{2t} -
\vec{p}_{t} \right) {\cal F}(x_1,k_{1t}^2,\mu^2) {\cal
F}(x_2,k_{2t}^2,\mu^2). \label{g_kt_factorization}
\end{eqnarray}

Here the four-momenta of the initial-state gluons are parameterized
as a sum of longitudinal and transverse parts
  $k_{1,2}=x_{1,2}P_{1,2}+k_{t1,2}$, $k_{t1,2}=(0,{\vec k}_{t1,2},0)$, $k_{1,2}^2=-{\vec k}_{t1,2}^2$, $P_{1,2}$
  are the four-momenta of the protons, $2P_1P_2=S$, $\overline{|{\cal M}_{R R \rightarrow g, gg, c \bar c }|^2}$ are the partonic
  cross sections with Reggeized gluons in the initial state.

Fully gauge invariant treatment of the initial-state off-shell
gluons can be achieved in $k_t$-factorization approach only when
they are considered as Reggeized gluons or Reggeons. The relevant
Reggeized amplitudes can be presented using Fadin-Kuraev-Lipatov
effective vertices: $C_{RR}^{g,\mu}$, $C_{RR}^{gg,\mu\nu}$ and
$C_{RR}^{q\bar q}$ \cite{FKL}. The squared amplitude of the partonic
subprocess $RR \to g$ is very simple and can be presented as
\begin{equation}
\overline{|{\cal M}_{R R \rightarrow g }|^2}=\frac{3}{2}\pi\alpha_S
{\vec p_t}^2.
\end{equation}
The useful analytical formulae for $\overline{|{\cal M}_{R R
\rightarrow g g }|^2}$ and $\overline{|{\cal M}_{R R \rightarrow c
\bar c }|^2}$ squared amplitudes are more complicated and we use the
ones as they have been written in Ref.~\cite{Nefedov:2013ywa}.

In the approach used here, the gluon unintegrated parton
distribution function (unPDF) ${\cal F}(x, k_{t}^2,\mu^2)$ is
normalized with respect to the collinear parton distribution function (PDF) by the following condition
  \begin{equation*}
  \int\limits^{\mu^2} d{k_t^2} {\cal F}(x,k_t^2,\mu^2) = x {\cal G}(x,\mu^2). \label{eq:norm_cond}
  \end{equation*}
A few phenomenological schemes to compute unPDFs of a proton where
proposed. In the present paper we use the LO
Kimber-Martin-Ryskin (KMR) unPDFs~\cite{KMR},
generated from the LO set of a up-to-date Martin-Motylinski-Harland-Lang-Thorne (MMHT2014) collinear PDFs~\cite{Harland-Lang:2014zoa} fitted also to the LHC data.

In the perturbative part of the calculations here we use a running LO $\alpha_{S}$
provided with the MMHT2014 PDFs. The charm quark mass used in the numerical calculations is $m_{c} = 1.5$ GeV. We set both the renormalization and factorization scales equal
to $\mu^{2} = p_{t}^{2}$ for $RR \to g$ subprocess, to the averaged transverse momentum $\mu^{2} = (p_{1t}^{2}+p_{2t}^{2})/2$ for $RR \to g g$, and to the averaged transverse mass $\mu^{2} = (m_{1t}^{2}+m_{2t}^{2})/2$ for $RR \to c \bar c$ case, where $m_{t} = \sqrt{p_{t}^{2} + m_{c}^{2}}$.

In order to calculate correlation observables for two mesons we
follow here, similar as in the single meson case, the fragmentation
function technique for hadronization process:
\begin{eqnarray}
\frac{d \sigma^{DPS}(pp \to D D X)}{d y_1 d y_{2} d^2 p_{1t}^{D} d^2 p_{2t}^{D}}
 &=&
\int \frac{D_{c \to D}(z_{1},\mu)}{z_{1}}\cdot \frac{D_{c \to D}(z_{2},\mu)}{z_{2}}\cdot
\frac{d \sigma^{DPS}(pp \to c c X)}{d y_1 d y_{2} d^2
  p_{1t}^{c} d^2 p_{2t}^{c}} d z_{1} d z_{2} \nonumber \\
  &+& \int \frac{D_{g \to D}(z_{1},\mu)}{z_{1}}\cdot \frac{D_{g \to D}(z_{2},\mu)}{z_{2}}\cdot
\frac{d \sigma^{DPS}(pp \to g g X)}{d y_1 d y_{2} d^2
  p_{1t}^{g} d^2 p_{2t}^{g}} d z_{1} d z_{2}\nonumber \\
  &+& \int \frac{D_{g \to D}(z_{1},\mu)}{z_{1}}\cdot \frac{D_{c \to D}(z_{2},\mu)}{z_{2}}\cdot
\frac{d \sigma^{DPS}(pp \to g c X)}{d y_1 d y_{2} d^2
  p_{1t}^{g} d^2 p_{2t}^{c}} d z_{1} d z_{2}, \nonumber \\
\end{eqnarray}
where: $p_{1t}^{g,c} = \frac{p_{1,t}^{D}}{z_{1}}$, $p_{2,t}^{g,c} =
  \frac{p_{2t}^{D}}{z_{2}}$ and meson momentum fractions $z_{1}, z_{2}\in (0,1)$.

The same formula for SPS $DD$-production via digluon fragmentation
reads
\begin{equation}
\frac{d \sigma^{SPS}_{gg}(pp \to D D X)}{d y_1 d y_{2} d^2 p_{1t}^{D} d^2 p_{2t}^{D}}
 \approx
\int \frac{D_{g \to D}(z_{1},\mu)}{z_{1}}\cdot \frac{D_{g \to D}(z_{2},\mu)}{z_{2}}\cdot
\frac{d \sigma^{SPS}(pp \to g g X)}{d y_1 d y_{2} d^2
  p_{1t}^{g} d^2 p_{2t}^{g}} d z_{1} d z_{2} \; ,
\end{equation}
where:
$p_{1t}^{g} = \frac{p_{1,t}^{D}}{z_{1}}$, $p_{2,t}^{g} =
  \frac{p_{2t}^{D}}{z_{2}}$ and
meson momentum fractions  $z_{1}, z_{2}\in (0,1)$.

In the case of calculations with the scale-independent Peterson FF the parameter $\varepsilon_{c} = 0.5$ is taken, which is the averaged value extracted from different $e^{+}e^{-}$ experiments and is commonly used in the literature. In turn, for predictions with the KKKS08 FF the evolution scale is set to the charm quark transverse mass $\mu^{2} = m_{t}^{2}$ and gluon transverse momentum $\mu^{2} = p_{t}^{2}$ for the $c \to D$ and $g \to D$ components, respectively.

In $e^{+}e^{-}$ collisions it is assumed naturally that gluons do not fragment to $D$ mesons for scales smaller than $\mu^{2}= \hat s = 4 m_{c}^{2}$.
In our calculation we have also tried to take $\mu^{2}= \hat s$ as the hadronization scale, which is an alternative to the typical choice of $\mu^{2} = m_{t}^{2}$ . In the first case naturally such initial scale is $4 m_{c}^{2}$ while for the second case the initial scale is set to $m_{c}^{2}$ so the effect of the evolution of fragmentation functions is present 
all over the phase space, however is very small for small transverse momenta of gluons.
In summary, both the choices of the hadronization scale lead to fairly similar final results for $DD$ correlations, which we have checked numerically. 

\section{Comparison to the LHCb data}

\begin{figure}[!h]
\begin{minipage}{0.47\textwidth}
 \centerline{\includegraphics[width=1.0\textwidth]{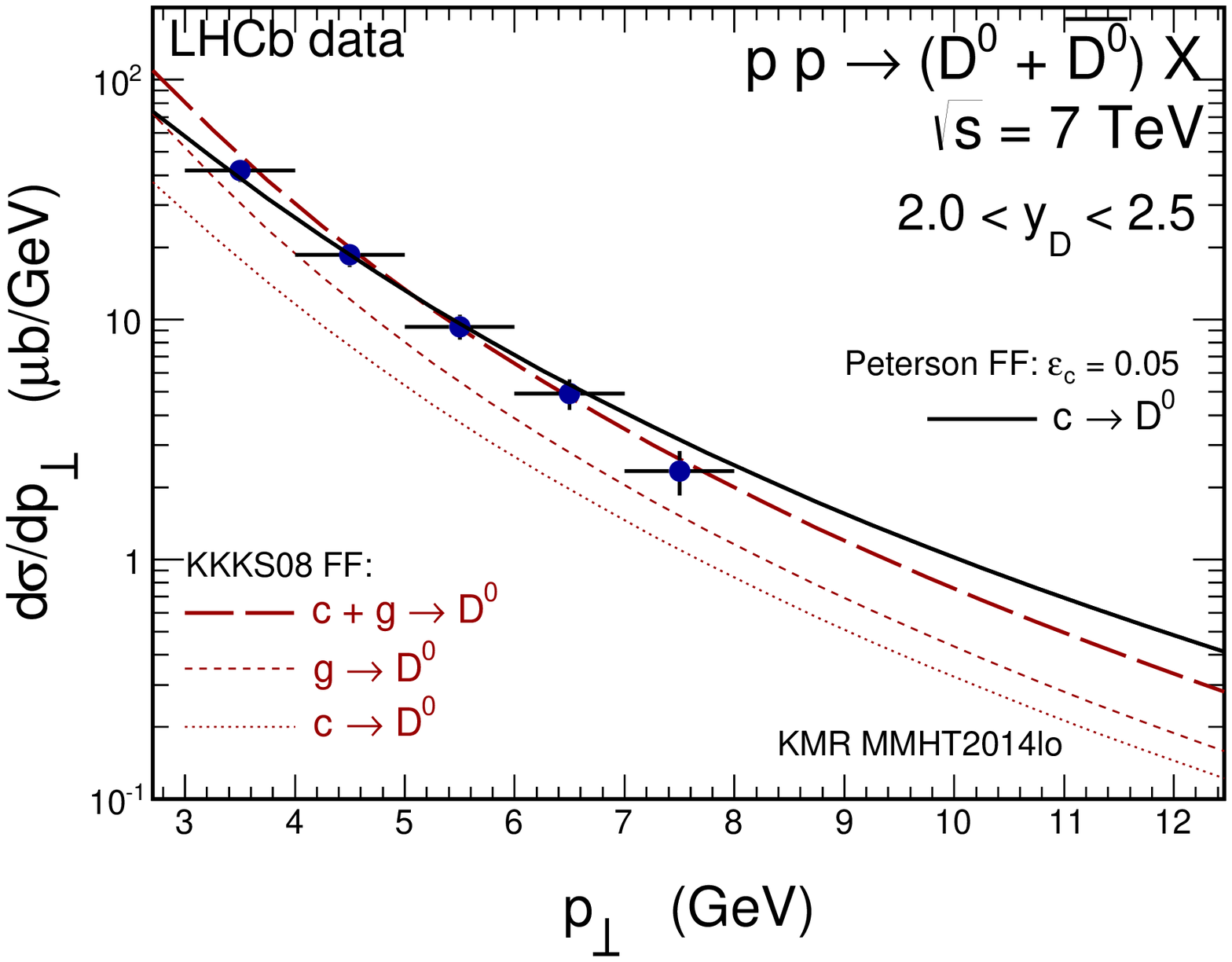}}
\end{minipage}
\hspace{0.5cm}
\begin{minipage}{0.47\textwidth}
 \centerline{\includegraphics[width=1.0\textwidth]{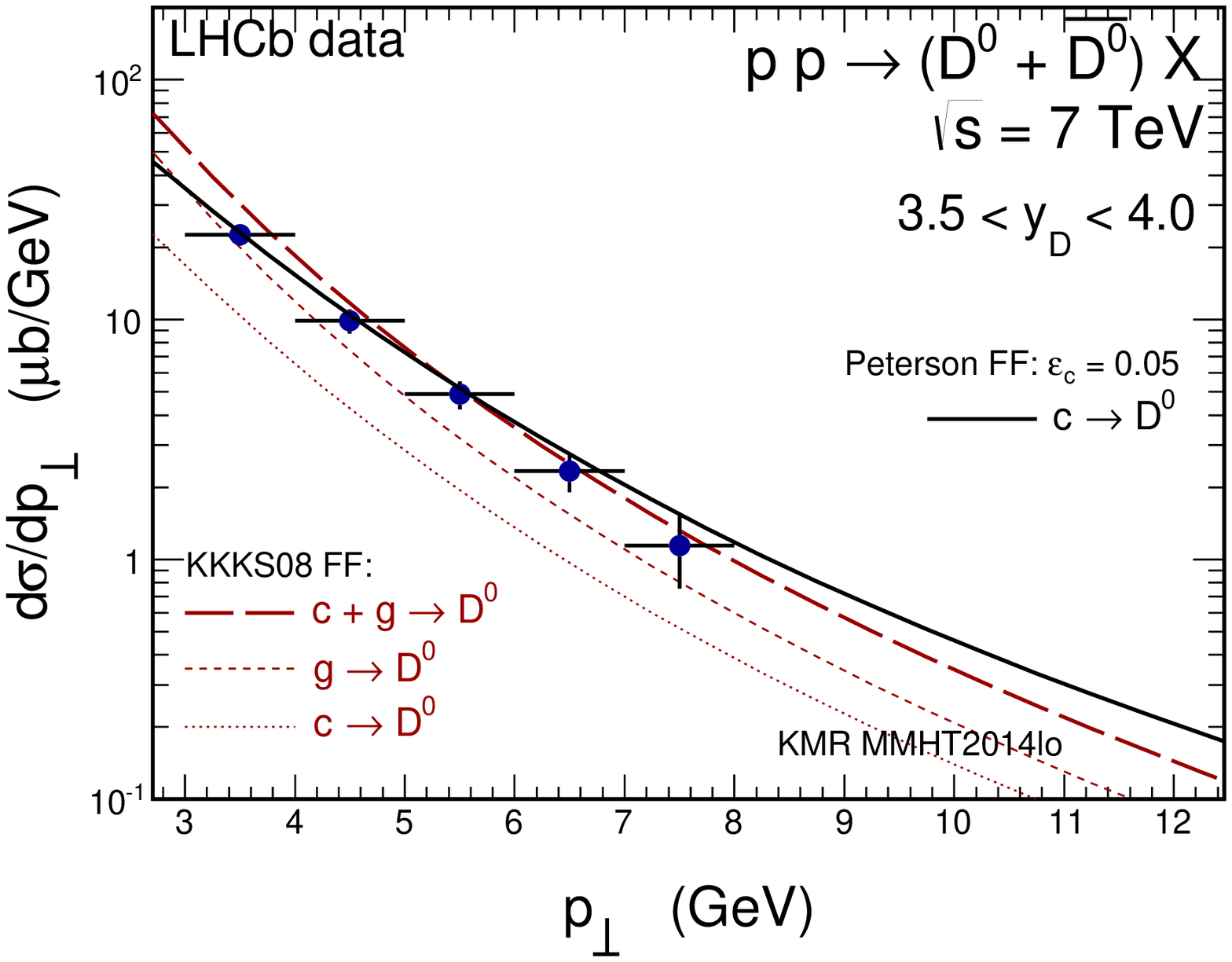}}
\end{minipage}
   \caption{
\small Charm meson transverse momentum distribution within the LHCb acceptance for inclusive single $D^{0}$ mesons (plus their conjugates) production.
Left and right panels correspond to two different rapidity intervals. 
Theoretical predictions for the Peterson $c \to D$ fragmentation function (solid lines)
are compared to the second scenario calculations with the KKKS08 fragmentation functions (long-dashed lines)
with $c \to D$ (dotted) and $g \to D$ (short-dashed) components
that undergo DGLAP evolution equation.
 }
 \label{fig:pTsingle}
\end{figure}

We start the presentation of our new results with a revision of inclusive single $D^{0}$ meson production measured some time ago by the LHCb collaboration \cite{Aaij:2013mga}. We already performed corresponding theoretical studies of the inclusive LHCb charm data based on both, the first (only $c \to D$) \cite{Maciula:2013wg} and the second ($c + g \to D$) scenario \cite{Nefedov:2014qea} in two separate papers. However, a direct comparison of the theoretical predictions based on these two scenarios for single $D$ meson production, calculated with the same set of $\alpha_{S}$, scales, unPDFs and other details, can be helpful for drawing definite conclusions in the following discussion of double $D$ meson production. As shown
in Fig.~\ref{fig:pTsingle}, both prescriptions give a very good description of the LHCb experimental data. Some small differences between them can be observed
for both very small and large meson transverse momenta. The latter effect can be recognized as a result of the DGLAP evolution which makes the slope
of the transverse momentum distribution in the second scenario a bit steeper than in the case of the first scenario, which is more favourable by the experimental data points.
In the region of very small $p_t$'s the second scenario gives larger cross sections and
slightly overestimates the experimental data points. This may come from the $g \to D$ fragmentation component which approaches a problematic region where $p_t \sim 2 m_{c}$. Then the treatment of charm quarks as massless in the DGLAP evolution of fragmentation function for very small evolution scale can be a bit questionable and may lead to a small overestimation of the integrated cross sections (especially in the case of $RR \to g \to D$ mechanism). 
We will come back to possible consequences of this effect when discussing $DD$ correlation observables.

\begin{figure}[!h]
\begin{minipage}{0.47\textwidth}
 \centerline{\includegraphics[width=1.0\textwidth]{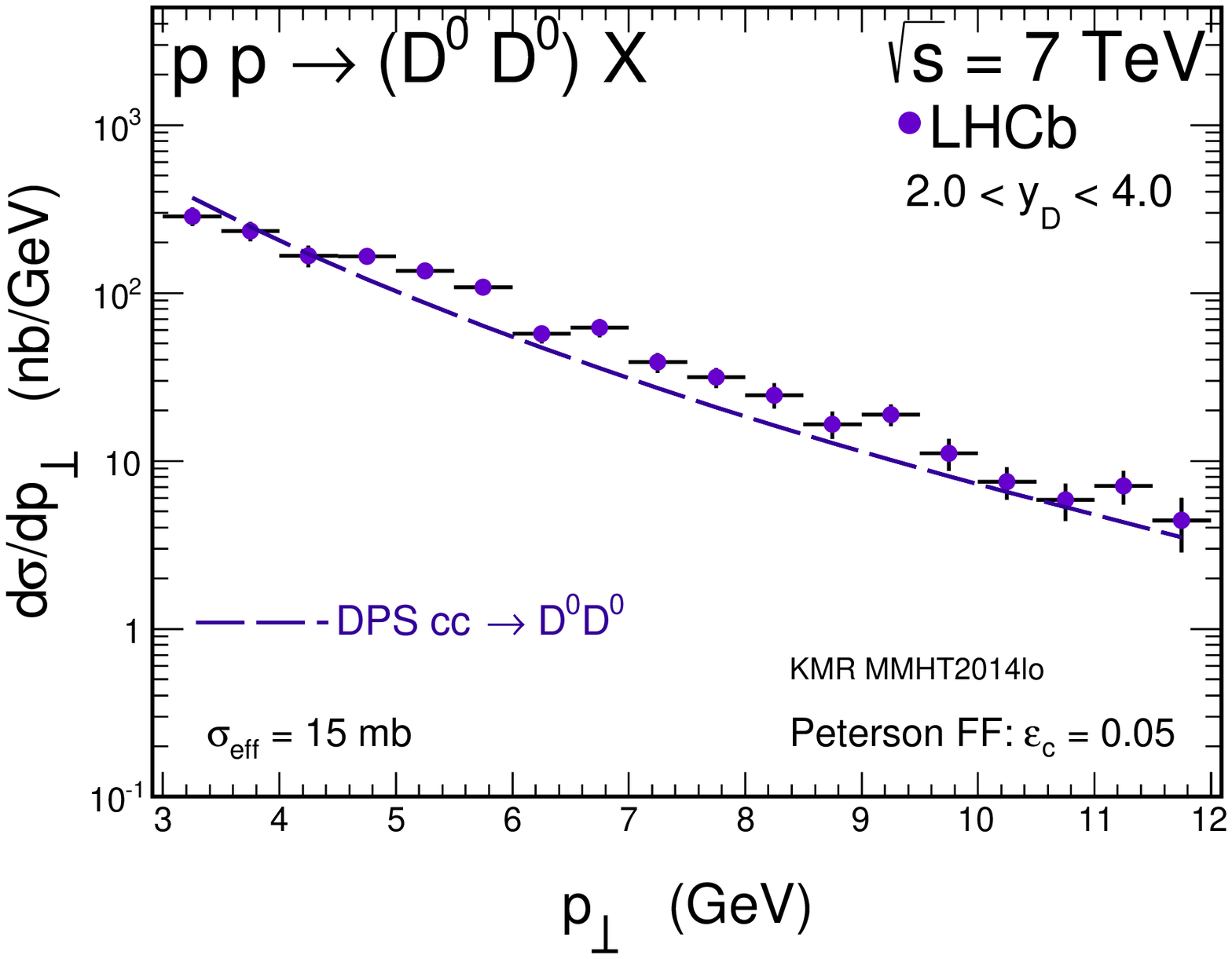}}
\end{minipage}
\hspace{0.5cm}
\begin{minipage}{0.47\textwidth}
 \centerline{\includegraphics[width=1.0\textwidth]{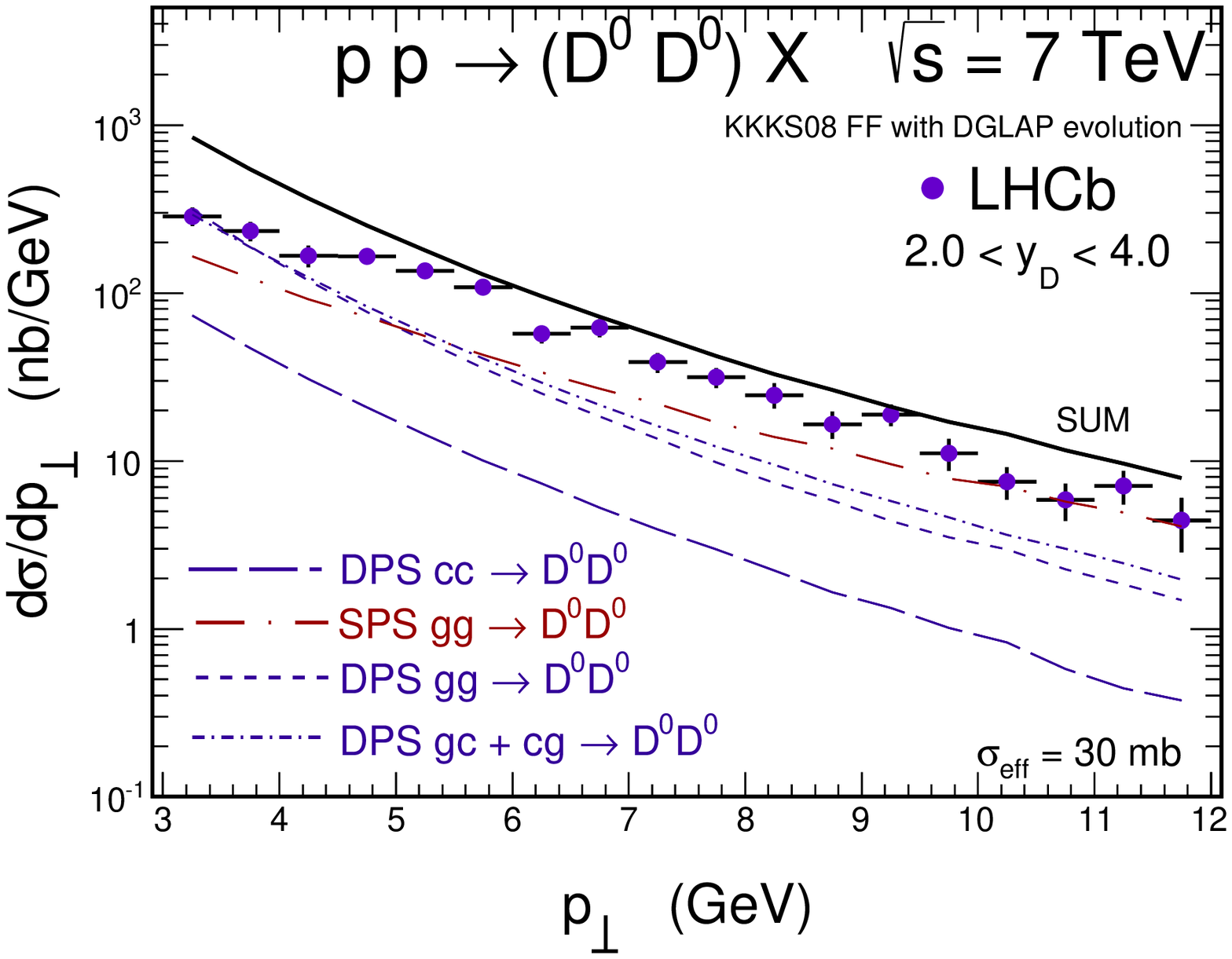}}
\end{minipage}
   \caption{
\small $D^0$ meson transverse momentum distribution within the LHCb
acceptance region.
The left panel is for the first scenario and for Peterson $c \to D$ fragmentation function
while the right panel is for the second scenario and for the fragmentation function
that undergo DGLAP evolution equation.
 }
 \label{fig:pT}
\end{figure}

Now we wish to compare results of our theoretical approach for double charm production described
briefly in the previous section with the LHCb experimental data for $D^{0}D^{0}$ pair production.
In Fig.~\ref{fig:pT} we compare results of our calculation
with experimental distribution in transverse momentum of one of the
meson from the $D^{0}D^{0}$ (or $\bar{D}^{0} \bar{D}^{0}$) pair. We show results for the first scenario when standard Peterson FF is used for the $c \to D^0$
(or $\bar c \to {\bar D}^0$) fragmentation (left panel) as well as
the result for the second scenario when the KKKS08 FFs with DGLAP evolution for $c \to D^0$ (or $\bar c \to \bar{D}^0$) and $g \to D^0$ (or $g \to \bar{D}^0$) are used. The results are almost independent of the scale of the fragmentation
function. The dependence on factorization scale of parton distributions
and on renormalization scale was discussed e.g. in Ref.~\cite{Maciula:2013kd}.
One can observe that the DPS $cc \to D^{0}D^{0}$ contribution in the new scenario is
much smaller than in the old scenario. In addition, the slope of the
distribution in transverse momentum changes. Both the effects are due to evolution of
corresponding fragmentation function in the second scenario, compared to
lack of such an effect in the first scenario.
The different new mechanisms shown in Fig.~\ref{fig:diagrams}
give contributions of similar size. We can obtain  agreement in the second case provided $\sigma_{eff}$ parameter is increased from conventional $15$ mb to $30$ mb. Even then we overestimate the LHCb data for $3 < p_{T} < 5$ GeV.

Can the increased value of $\sigma_{eff}$ = 30 mb be understood?
First of all the LHCb experiment measures charmed mesons (charm
quarks/antiquarks) in forward directions. As shown in 
Ref.~\cite{Gaunt:2014rua} at larger charm quark/antiquark rapidities
the relative contribution of perturbative partonic splitting increases.
The $\sigma_{eff}$ parameter includes both conventional $2v2$ uncorrelated 
and correlated single parton $2v1$ splitting contribution.
As shown in Ref.~\cite{Gaunt:2014rua} the smaller perturbative single parton splitting
contribution the larger $\sigma_{eff}$.
Also the conventional uncorrelated parton picture may be too simplistic.
The nonperturbative correlations may lead to the effective dependence
of $\sigma_{eff}$ on $c$ and/or $\bar c$ transverse momentum 
(see a recent model analysis for jet production in Ref.~\cite{Ostapchenko:2015vmy}). 

\begin{figure}[!h]
\begin{minipage}{0.47\textwidth}
 \centerline{\includegraphics[width=1.0\textwidth]{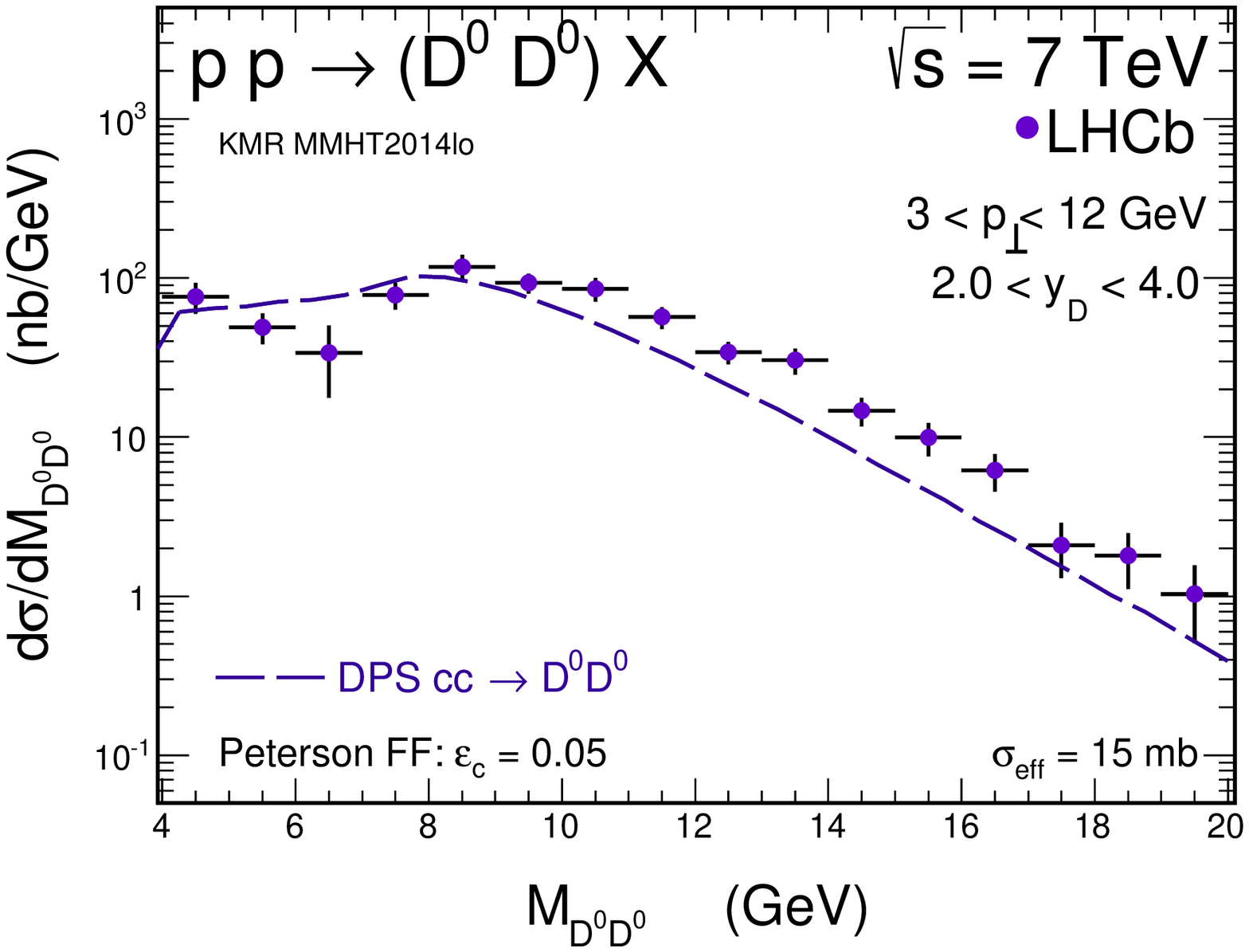}}
\end{minipage}
\hspace{0.5cm}
\begin{minipage}{0.47\textwidth}
 \centerline{\includegraphics[width=1.0\textwidth]{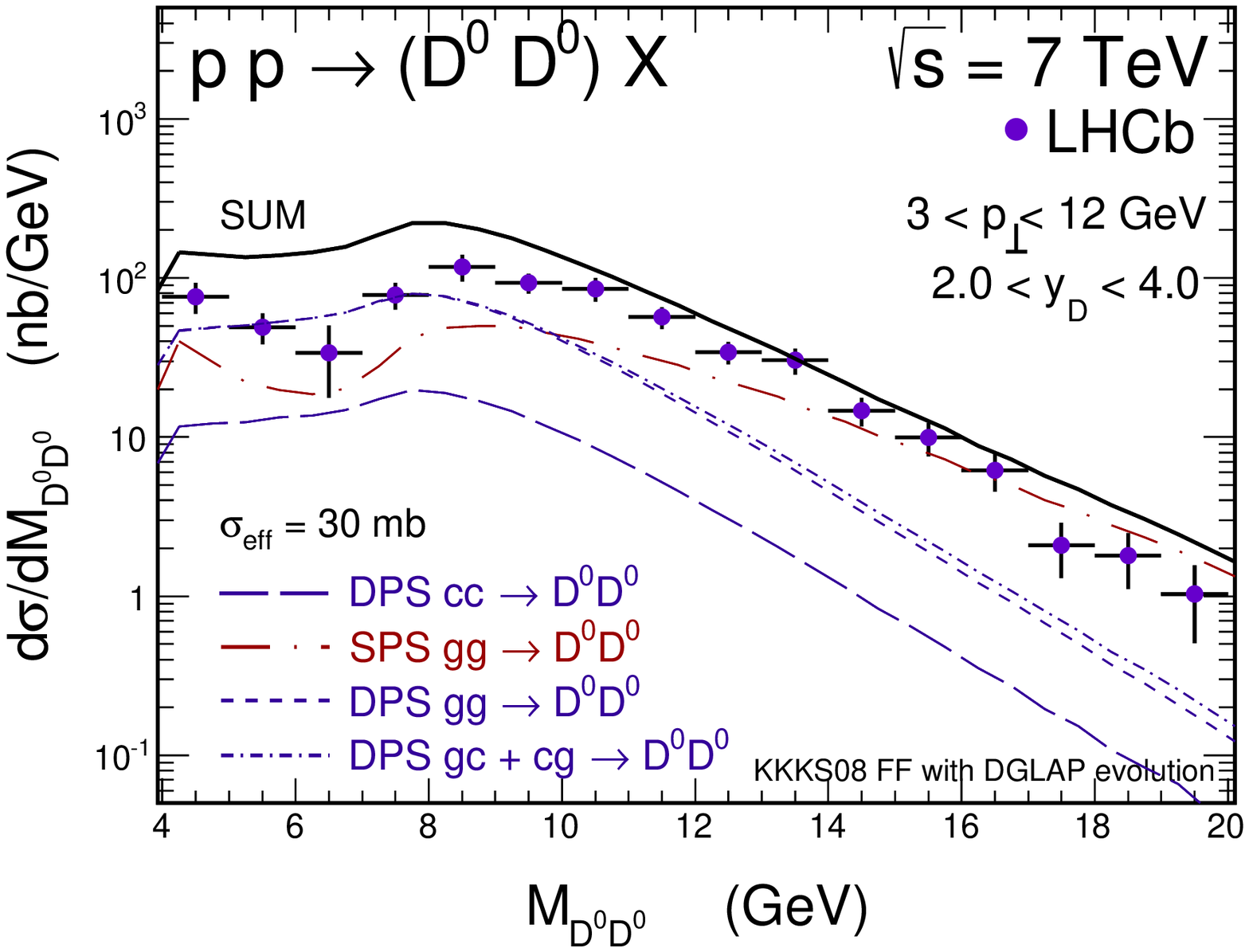}}
\end{minipage}
   \caption{
\small $M_{D^0 D^0}$ dimeson invariant mass distribution within the LHCb acceptance
region.
The left panel is for the first scenario and for the Peterson $c \to D$ fragmentation function
while the right panel is for the second scenario and for the fragmentation function
that undergo DGLAP evolution equation.
 }
 \label{fig:Minv}
\end{figure}

In Fig.~\ref{fig:Minv} we show dimeson invariant mass distribution $M_{D^0 D^0}$
again for the two cases considered. In the first scenario we get a good agreement only for small invariant 
masses while in the second scenario we get a good agreement
only for large invariant masses. The large invariant masses are
strongly correlated with large transverse momenta, so the situation 
here (for the invariant mass distribution) is quite similar as 
in Fig.~\ref{fig:pT} for the transverse momentum distribution.

\begin{figure}[!h]
\begin{minipage}{0.47\textwidth}
 \centerline{\includegraphics[width=1.0\textwidth]{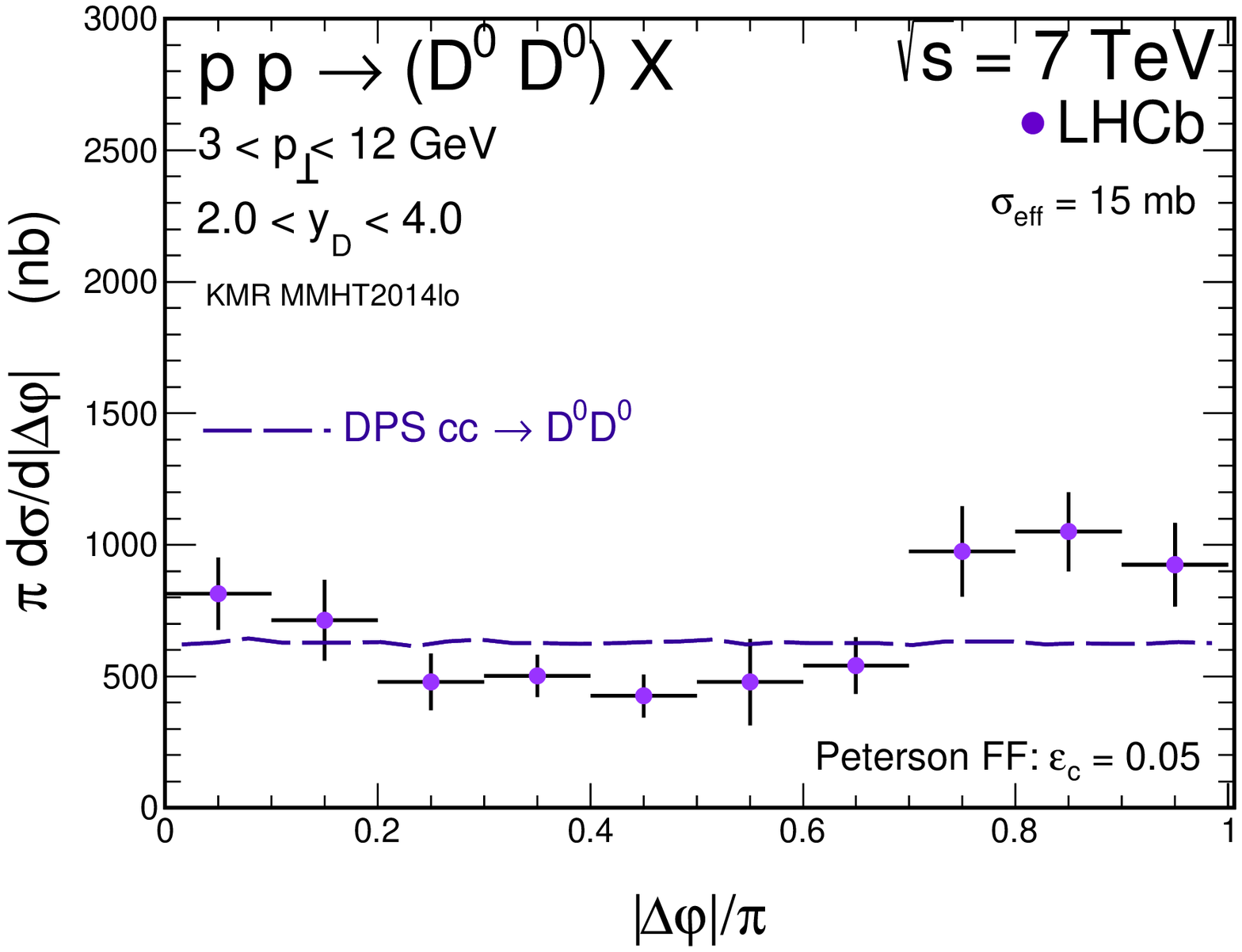}}
\end{minipage}
\hspace{0.5cm}
\begin{minipage}{0.47\textwidth}
 \centerline{\includegraphics[width=1.0\textwidth]{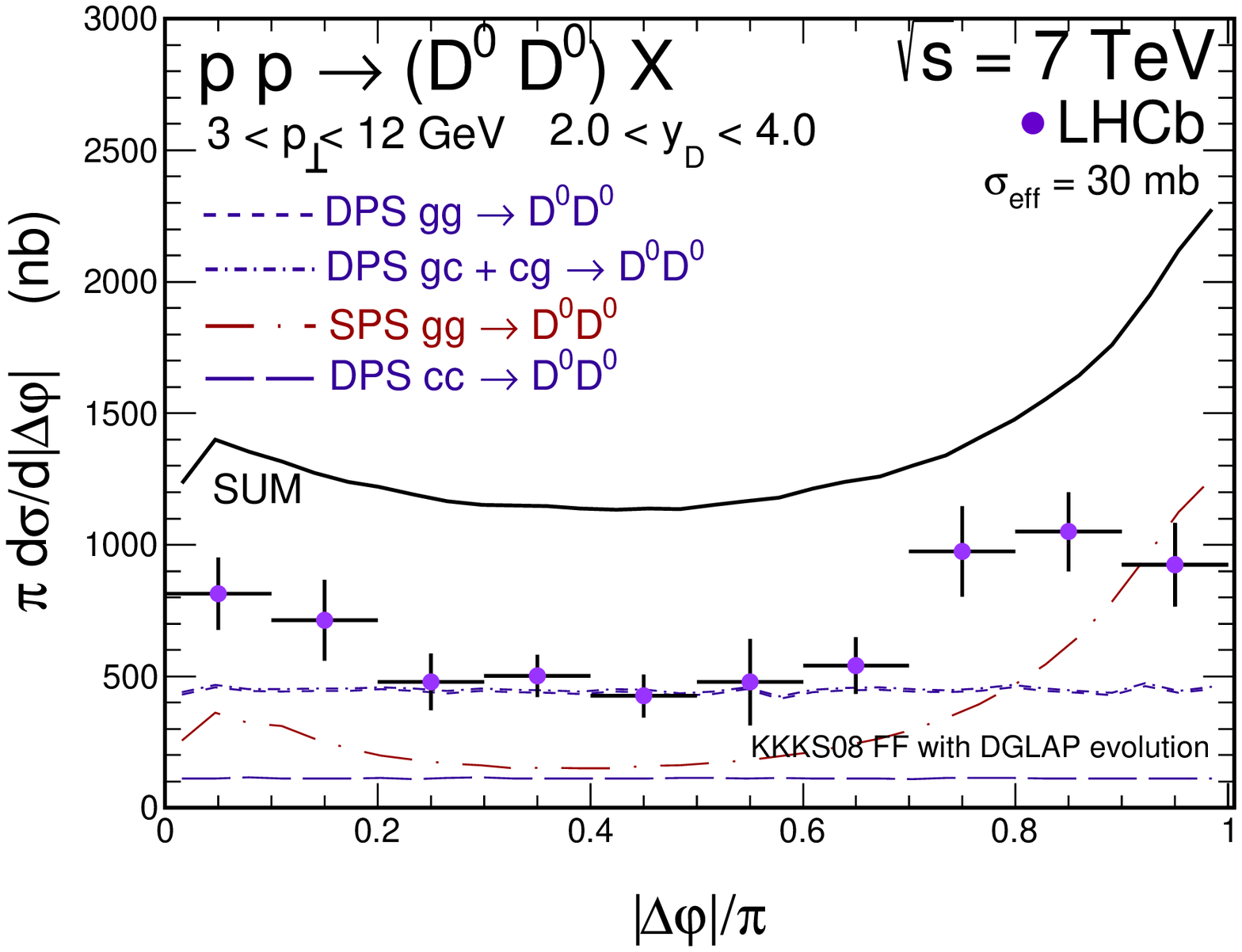}}
\end{minipage}
   \caption{
\small Distribution in azimuthal angle $\varphi_{D^0 D^0}$ between the two $D^0$ mesons
within the LHCb acceptance region.
The left panel is for the first scenario and for Peterson $c \to D$ fragmentation function
while the right panel is for the second scenario and for the fragmentation function
that undergo DGLAP evolution equation.
 }
 \label{fig:phid}
\end{figure}

In Fig.~\ref{fig:phid} we show azimuthal angle correlation $\varphi_{D^0 D^0}$ 
between $D^0$ and $D^0$ (or ${\bar D}^0$ and ${\bar D}^0$ mesons).
While the correlation function in the first scenario is completely flat,
the correlation function in the second scenario shows some tendency similar as in the experimental data.
The increase at small $\Delta \varphi$ for the SPS $gg \to D^0 D^0$
contribution is due to s-channel pole in the amplitude for $RR \to gg$
which we regularize by $\hat{s} > 4 m_{c}^{2}$ condition. In the
$k_t$-factorization, initial partons have transverse momenta, but final
gluons may have equal rapidities even when $\Delta \varphi$ is far from $\pi$.
The observed overestimation comes from the region of small transverse momenta.
The situation may be improved when a proper transverse momentum dependence of $\sigma_{eff}$
will be included, but this needs further studies.

\begin{figure}[!h]
\begin{minipage}{0.47\textwidth}
 \centerline{\includegraphics[width=1.0\textwidth]{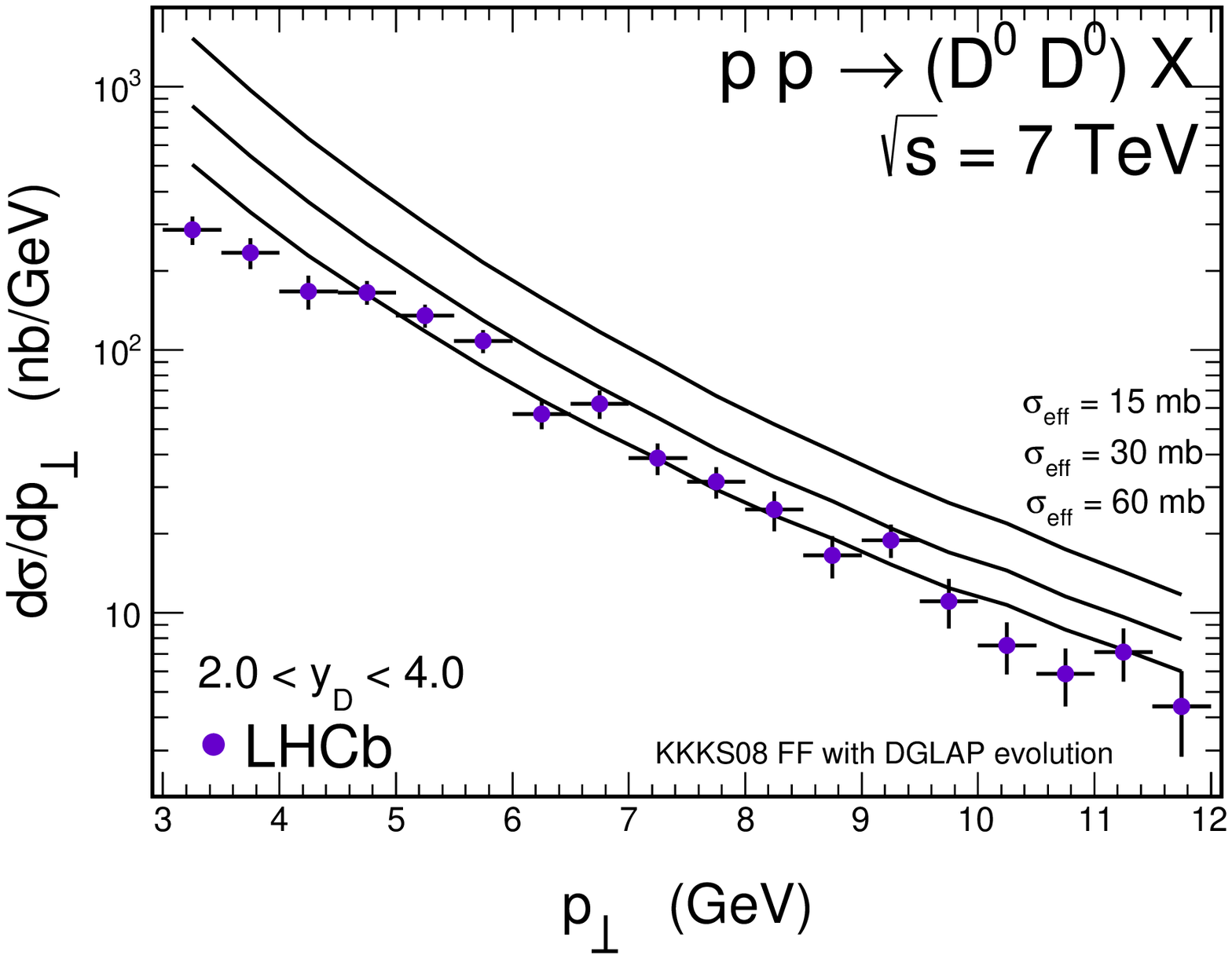}}
\end{minipage}
\hspace{0.5cm}
\begin{minipage}{0.47\textwidth}
 \centerline{\includegraphics[width=1.0\textwidth]{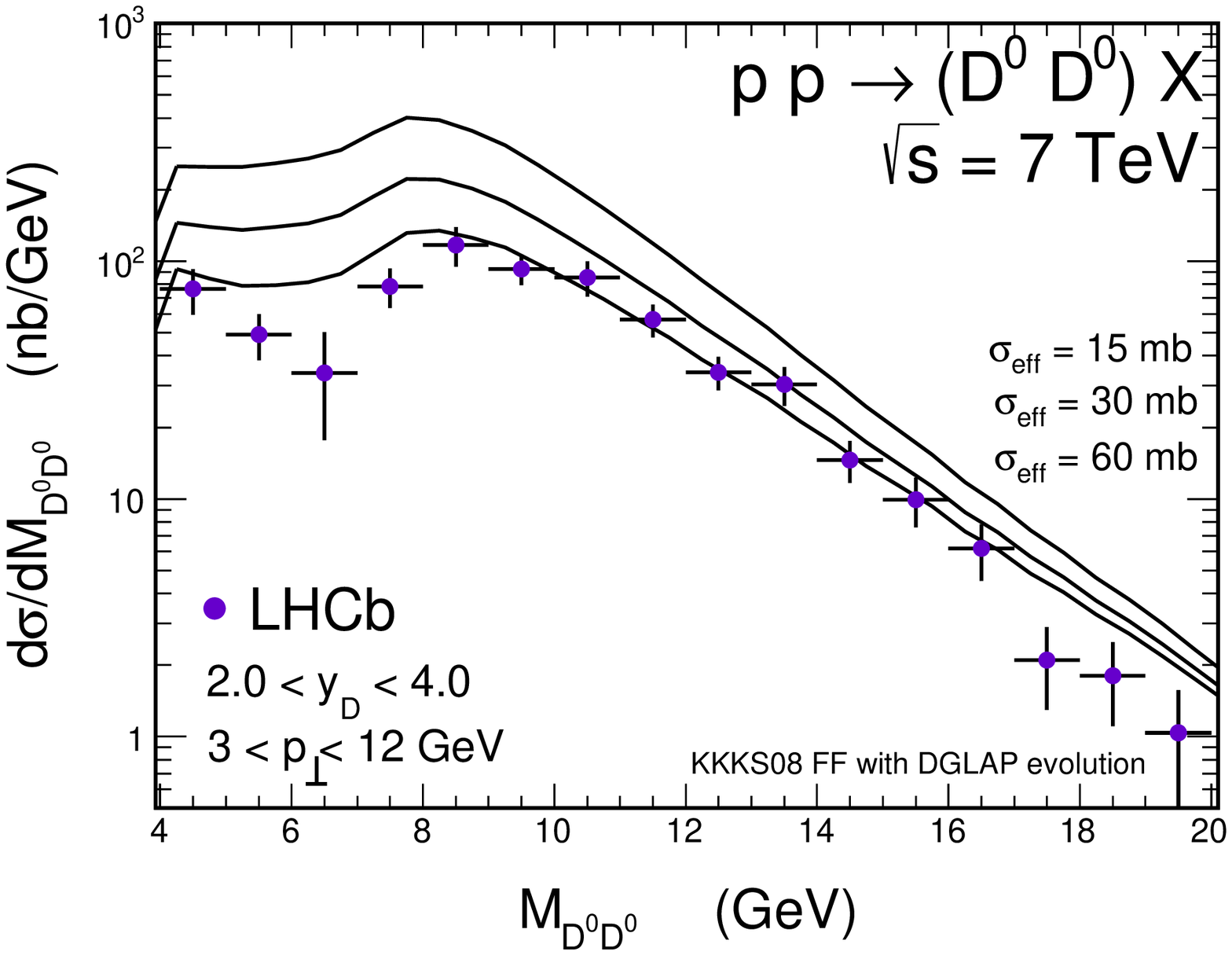}}
\end{minipage}\\
\begin{minipage}{0.47\textwidth}
 \centerline{\includegraphics[width=1.0\textwidth]{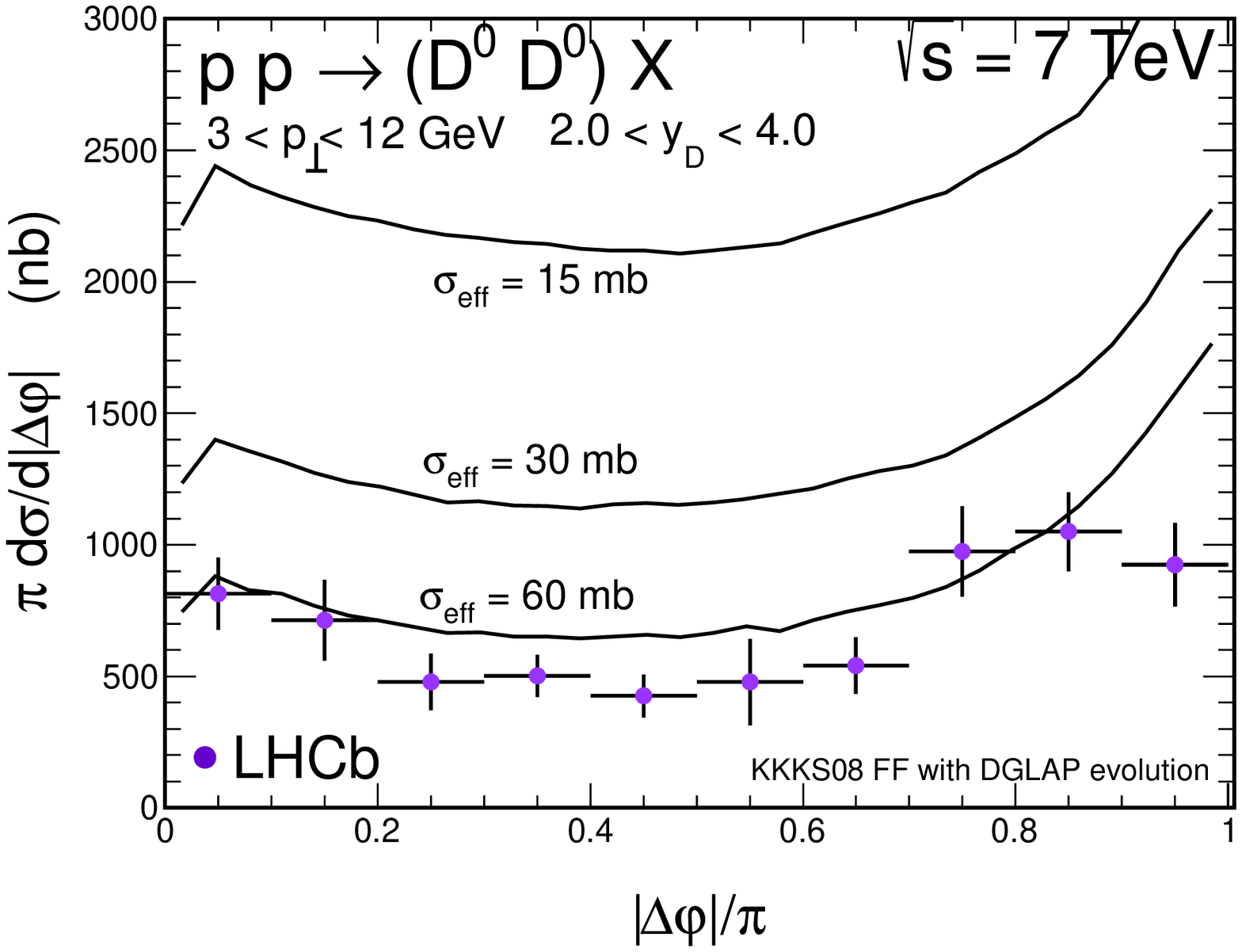}}
\end{minipage}
   \caption{
\small The dependence of the results of the second scenario on the parameter $\sigma_{eff}$ used in the calculation of the DPS contributions.
Here the three lines correspond to $\sigma_{eff}$ equal to $15$, $30$, and $60$ mb, from top to bottom, respectively.}
 \label{fig:sig_eff}
\end{figure}

Finally we wish to summarize the present situation for the second scenario. 
In Fig.~\ref{fig:sig_eff} we show the different distributions discussed above for different values of $\sigma_{eff}$.
Good description can be obtained only for extremely large values of $\sigma_{eff}$ which
goes far beyond the geometrical picture \cite{Gaunt:2014rua} and that are much larger than for other reactions
and in this sense is inconsistent with the factorized Ansatz. We think that the solution
of the inconsistency is not only in the DPS sector as already discussed in this paper.

\section{Conclusions}

In the present paper we have discussed production of $D^0D^0$ or
${\bar D}^0{\bar D}^0$ meson-meson pairs in proton-proton collisions at the LHC. We have considered
the double-parton scattering mechanism of double $c \bar c$ production
and subsequent double hadronization of two $c$ quarks or two $\bar c$
antiquarks using $c \to D^0$ or $c \to {\bar D}^0$ fragmentation
functions that undergo DGLAP evolution equation with one of the traditional,
scale-independent fragmentation function used as an input at the initial scale that is set to $\mu^2 =$ 4 $m_c^2$.

In addition we have included also production of gluonic dijets and their
subsequent hadronization to the neutral pseudoscalar $D$ mesons.
The $g \to D$ fragmentation function is assumed to be zero at the
initial scale that is set to $\mu^2 =$ 4 $m_c^2$.
Also mixed $g \to D$ and $c \to D$ mechanisms occur naturally.

We find that at $\sqrt{s}$ = 7 TeV the two mechanisms give similar
contribution for the LHCb experimental acceptance.
While the DPS mechanism dominates at small $D$ meson transverse momenta,
the SPS double gluon fragmentation takes over for larger transverse momenta.

When added together the new mechanisms give similar result as the first scenario
with one subprocess ($cc \to DD$) and fixed (scale-independent) fragmentation function.
However, some correlation observables, such as dimeson invariant mass
or azimuthal correlations between $D$ mesons, are slightly better
described.

In our calculation within the second scenario a larger value 
of $\sigma_{eff}$ is needed to describe the LHCb data than found from the review of several 
experimental studies of different processes.
This can be partially understood by a lower contribution of
perturbative parton splitting as found in Ref.~\cite{Gaunt:2014rua}
and/or due to nonperturbative correlations in the nucleon which may lead
to transverse momentum dependent $\sigma_{eff}$.
Clearly more involved studies are needed to understand the situation
in details. Some problem may be also related to the fact
that the fragmentation function used in the second scenario
were obtained in the DGLAP formalism with massless $c$ quarks and 
$\bar c$ antiquarks which may be a too severe approximation,
especially for low factorization scales (i.e. low transverse momenta)
for fragmentation functions.
We expect that including mass effects in the evolution would lower
the $g \to c$ (or $g \to \bar c$) fragmentation. Such a study would be
useful but clearly goes beyond the scope of the present paper.
At present one may only expect that the final (fully consistent) result 
should be in between the old and new (not completely consistent at present) 
approach.

In this context we remind a trial to describe the correlation
observables in more involved non-factorized approach to DPS
\cite{Echevarria:2015ufa}. The authors there neglected hadronization
and worked in leading-order collinear approach.
However, they were not able to describe the details of the
LHCb distributions.

The presence of the new SPS mechanism may mean that the extraction
of $\sigma_{eff}$ directly from the LHCb experimental data \cite{Aaij:2012dz}
may be not correct.

We expect that at higher energies (for example for Future Circular Collider) the proportions will change and
at asymptotically high energies, much above the LHC energies, the DPS
mechanism will win.

\vspace{1cm}

{\bf Acknowledgments}

We are particularly indebted to Anton Karpishkov for discussion of several technical issues
and Ingo Schienbein for explaining some details of KKKS08 fragmentation funkctions.
This study was partially
supported by the Polish National Science Center grant
DEC-2014/15/B/ST2/02528 and by the Center for Innovation and
Transfer of Natural Sciences and Engineering Knowledge in
Rzesz{\'o}w. The work was supported by Russian Foundation for Basic
Research through the Grant No~14-02-00021 and by the Ministry of
Education and Science of Russia under Competitiveness Enhancement
Program of SSAU for 2013-2020.

\end{document}